\documentclass{elsart4}

\usepackage{amssymb}
\usepackage{graphicx}

\usepackage[english,francais]{babel}

\newtheorem{e-proposition}[theorem]{Proposition}

\newtheorem{e-definition}[theorem]{Definition\rm}

\renewcommand{\theequation}{\arabic{equation}}
\def\og{\leavevmode\raise.3ex\hbox{$\scriptscriptstyle\langle\!\langle$~}}
\def\fg{\leavevmode\raise.3ex\hbox{~$\!\scriptscriptstyle\,\rangle\!\rangle$}}

\begin{document}
\centerline{The Chinese-French {\it SVOM} mission for gamma-ray burst studies}
\begin{frontmatter}


\selectlanguage{english}
\title{The Chinese-French {\it SVOM} mission for gamma-ray burst studies}


\selectlanguage{english}
\author[paul]{Jacques Paul}
\ead{jacques.paul@cea.fr}
\author[wei]{Jianyan Wei}
\ead{wjy@bao.ac.cn}
\author[basa]{St\'ephane Basa}
\ead{stephane.basa@oamp.fr}
\author[zhang]{Shuang-Nan Zhang}
\ead{zhangsn@ihep.ac.cn}

\address[paul]{CEA Saclay, DSM/IRFU/Service d'Astrophysique, 91191, Gif-sur-Yvette, France and \\ Laboratoire d'Astroparticule et Cosmologie, 10, rue Alice Domon et L\'eonie Duquet, 75205 Paris, France}
\address[wei]{National Astronomical Observatories, Chinese Academy of Sciences, Beijing 100012, China}
\address[basa]{Laboratoire d'Astrophysique de Marseille, 38 rue Joliot-Curie, 13388 Marseille, France}
\address[zhang]{Laboratory of Particle Astrophysics, Institute of High Energy Physics, Chinese Academy of Sciences, Beijing 100049, China}

\begin{abstract}
We present the {\it SVOM} (Space-based multi-band astronomical Variable Objects Monitor) mission that the Chinese National Space Agency and the French Space Agency have decided to jointly implement. {\it SVOM} has been designed to detect all known types of gamma-ray bursts (GRBs), to provide fast and reliable GRB positions, to measure the broadband spectral shape and temporal properties of the GRB prompt emission, and to quickly identify the optical/near-infrared afterglows of detected GRBs, including high-redshift ones. Scheduled to be in orbit in the second half of the present decade, the {\it SVOM} satellite will carry a very innovative scientific payload combining for the first time a wide field X- and gamma-ray coded mask imager for GRB real-time localizations to few arcmin, a non-imaging gamma-ray monitor, and two narrow-field instruments for the study of the GRB early afterglow emission in the X-ray and visible bands. The {\it SVOM} payload is complemented by ground-based instruments including a wide-field camera to catch the GRB prompt emission in the visible band and two robotic telescopes to measure the photometric properties of the early afterglow. A particular attention is paid to the GRB follow-up in facilitating the observation of the {\it SVOM} detected GRB by the largest ground based telescopes.

\vskip 0.5\baselineskip

\selectlanguage{francais}
\noindent{\bf R\'esum\'e}
\vskip 0.5\baselineskip
\noindent
{\bf La mission sino-fran\c{c}aise SVOM pour l'\'etude des sursauts gamma. }
Nous pr\'esentons la mission SVOM dont l'Agence Nationale Chinoise de l'Espace et le Centre National dÕ\'Etudes Spatiales  ont d\'ecid\'e ensemble la r\'ealisation. SVOM (Moniteur spatial multi-longueurs d'onde d'objets variables) a \'et\'e con\c{c}u pour d\'etecter tous les types de sursauts gamma, pour en fournir promptement une position pr\'ecise, pour mesurer le spectre dans un large intervalle de fr\'equence  et les propri\'et\'es temporelles de leur \'emission prompte et pour identifier rapidement dans le visible ou le proche infrarouge leur \'emission r\'emanente, y compris celle des sursauts pr\'esentant de grands d\'ecalages vers le rouge. En orbite dans la deuxi\`eme moiti\'e de la d\'ecennie, le satellite SVOM embarquera une charge utile scientifique tr\`es innovante, combinant pour la premi\`ere fois une cam\'era \`a grand champ \`a ouverture cod\'ee pour localiser les sursauts gamma avec une pr\'ecision de quelques minutes d'arc dans la bande des rayons X et des rayons gamma de basse \'energie, un moniteur gamma \`a grand champ et deux instruments \`a petit champ pour \'etudier l'\'emission r\'emanente pr\'ecoce dans les domaines des rayons X et du visible. La charge utile embarqu\'ee \`a bord de SVOM est compl\'et\'ee par des instruments au sol dont une cam\'era \`a grand champ pour enregistrer dans le visible l'\'emission prompte des sursauts et deux t\'elescopes robotiques pour mesurer les propri\'et\'es photom\'etriques de leur r\'emanence pr\'ecoce. Le suivi des sursauts est l'objet d'une attention particuli\`ere en permettant aux plus grands t\'elescopes de la plan\`ete d'observer dans les meilleures conditions les sursauts d\'etect\'es par SVOM.

\keyword{SVOM; Gamma-ray bursts; Cosmology} \vskip 0.5\baselineskip
\noindent{\small{\it Mots-cl\'es~:} SVOM~; Sursaut gamma~; Cosmologie}}
\end{abstract}
\end{frontmatter}

\selectlanguage{english}

\section{Introduction}
It is only in the last decade of the past century that the full astrophysical potential of gamma-ray bursts (GRBs) has been gradually realized, with the discovery of their counterparts in the X-ray, visible and radio bands. They are extremely luminous transient sources appearing when a newborn stellar mass black hole or magnetar emits an ultra-relativistic jet towards the Earth. For several seconds or minutes, GRBs become the brightest sources of the Universe, illuminating their surrounding medium, and scanning the matter along their line-of-sight to the Earth. GRBs also involve remarkable physics, only found in few astronomical sites where a huge amount of energy is deposited in a small volume of space.

Consequently, the study of GRBs has the potential to expand our understanding of key astrophysical issues. In the next years, they will undoubtedly shed new light on the evolution of the young Universe, particularly on the history of star formation and the re-ionization of the intergalactic medium. In parallel, they will bring crucial new data on the radiation processes at work in regions of space containing a huge energy density and they will provide reliable alerts for gravitational wave and neutrino detectors. In order to fulfill these promises, future GRB studies must rely on the availability of a continuous flow of accurate positions (to take advantage of instrumental progress), but also on the measure of many additional parameters (redshift, E$_{peak}$, jet break time, ...), crucial for the understanding of the GRBs and for their use as astrophysical tools.

Aiming to contribute to the field of GRB research, the Chinese Space Agency (CNSA), the French Space Agency (CNES), and research laboratories from both countries are developing the {\it SVOM} mission (Space-based multi-band astronomical Variable Object Monitor). The {\it SVOM} mission, which has successfully concluded its feasibility study (phase A), and has started its detailed design phase for a launch in the second half of the present decade, has been designed to (i) permit the detection of all known types of GRBs, (ii) provide fast, reliable GRB positions, (iii) measure the broadband spectral shape of the prompt emission (from visible to MeV), (iv) measure the temporal properties of the prompt emission (from visible to MeV), (v) quickly identify the afterglows of detected GRBs at X-ray and optical/near-infrared wavelengths, including those with high redshifts, (vi) measure the broadband spectral shape of the early and late afterglow (from visible to X-rays), and (vii) measure the temporal evolution of the early and late afterglow (from visible to X-rays).

\section{Scientific rationale of the {\it SVOM} mission}
\subsection{Theory of GRBs}
It is now commonly accepted that GRBs are the manifestation of a sole phenomenon: the ejection of an ultra-relativistic jet by a new-born rapidly-spinning compact object (black-hole or magnetar) formed at the end of the evolution of massive stars and by mergers of compact objects. The ultra-relativistic motion of the jet strongly amplifies the light in the direction of the motion, making GRBs so luminous that they are visible across the entire Universe for observers close to the jet axis. But the sources of these jets, their nature, and the conditions to launch them are far from being understood.

GRBs are often explained with three radiating components: (i) internal shocks due to the variability of the central engine and which propagate within the outflow where they produce the prompt emission, (ii) forward shocks due to the deceleration by the surrounding medium where they propagate and where they produce the afterglow emission, radiating from X-ray to radio, and (iii) reverse shocks (due to the deceleration by the surrounding medium but propagating within the outflow) which may explain the visible and/or X-ray ashes observed during or just after the end of the prompt gamma-ray emission. Photospheric emission is also expected but not clearly identified. For more details, see the chapters by Zhang and by Godet \& Mochkovitch. 

This scenario explains the incredible luminosity of the prompt emission and its characteristic energy, the spectro-temporal evolution of the afterglow, the association of GRBs with type Ibc supernovae, the predominance of GRBs in star-forming host galaxies. However, this scenario cannot explain the extreme diversity of the spectro-temporal behavior of GRBs, which are being disclosed by prompt visible and X-ray observations triggered by {\it Swift}.

GRBs are particularly well suited to study the physics at work in relativistic jets. Progress in this field will require time-resolved panchromatic observations of GRBs performed as soon as possible after the burst and during the prompt emission whenever this is possible. The nature of the physical mechanism that launches GRB ultra-relativistic jets is not elucidated, and the nature of the jet (electromagnetic or matter dominated) is currently the subject of an intense debate. For example, the canonical behavior of X-ray afterglow light-curves that have been recently revealed by {\it Swift} seems inconsistent with their light-curves recorded in the visible in many cases, which poses a major challenge to the standard model.

\subsection{Toward a better understanding of the GRB phenomenon}
One of the major discoveries of the present generation of GRB missions is the diversity of the GRB population, which includes classical GRBs, X-ray flashes (XRFs), soft GRBs with little emission above 30 keV, short/hard GRBs, sub-luminous GRBs, optically dark GRBs. While the energy dissipation and radiation physics of GRBs may be rather similar, their progenitors may be quite diverse. Essentially two types of progenitors have been proposed: massive stars collapsing into a compact object (black-hole or magnetar) and mergers of compact objects. Convincing piece of evidence exists that links GRBs with the end of life of massive stars ($\ge$ 20 M$_\odot$). The strongest evidence is the fact that several nearby long GRBs are optically associated with Type Ibc supernovae. However, radio surveys suggest that less than one type Ibc supernova out of ten is associated with a GRB: their production must depend on some special conditions. Models suggest that the fast rotation of the stellar core of the progenitor is essential for the production of GRBs. The origin of this fast rotation and the mechanisms which make GRB progenitors different from other massive stars are still very poorly understood. For more details, see the chapter by Zhang.

We do not have convincing clues on the nature of the progenitors of short GRBs but there is strong suspicion that they could be due to mergers of compact objects. They are found in all types of galaxies, on the contrary to long GRBs that are exclusively found in star forming galaxies. This is the consequence of the long lifetime of the supposed progenitors of short GRBs: even if the progenitor is formed during a period of stellar formation, the long delay between its formation and the GRB allows this one to occur when the stellar formation has ended. The long life-time of short GRB progenitors supports the merger hypothesis since binary systems of compact objects need to lose their angular momentum by gravitational radiation before they can merge.

\subsection{GRBs and cosmology}
The detection of very distant bright GRB afterglows has opened a new window on our Universe. This breakthrough is due to the extreme luminosity of these events, which makes them powerful probes of the Universe at very high redshift. The most distant GRB observed until now, GRB 090423, has a redshift z $\approx$ 8.2, significantly greater than those of the most distant quasars. This property makes GRBs exceptional beacons to study the early Universe. In particular, they offer a unique possibility to study the epoch of the first star formation and of the re-ionization of the Universe. The first stars could be very massive (100 M$_\odot$ or more), a fraction of them could have produced detectable GRBs. For more details, see the chapter by Petitjean \& Vergani.

\underline{Cosmological lighthouses.} Like bright quasars, GRBs can be used as probes of cosmological lines-of-sight. Their advantages are many-fold. First, they allow probing galaxies selected thanks to their metal absorption up to the highest redshifts, where there might be no other observational techniques to find faint bound objects. Second, GRBs, as bright and transient background sources, will be ideal tools to observe extreme environments in absorbers. Many high column density systems have already been found at GRB positions, and it is expected that they will permit probing systems with large column density and/or dusty systems, which are missing from current quasar absorber surveys (because of the dust extinguishing the quasar). Third, the central question of the relation of these absorbers with galaxies remains. Indeed, the steadily bright quasars make it extremely difficult to search for the galaxy counterparts of the intervening absorbers. In GRBs, the afterglows will disappear and the search for emission from the absorbing galaxy will only be affected by the light from the much fainter GRB host galaxy. Absorbers detected towards GRBs will therefore be the perfect tool to bridge systems detected in absorption with galaxies detected in emission.

\underline{Host galaxies.} GRBs are found in their vast majority in blue irregular galaxies which do not contribute to the bulk of star formation. In addition, the GRB-host population seems biased with respect to the local population of the hosts of core collapse supernovae. This means that all massive stars do not contribute equally to the production of GRBs. One suggestion is that only the massive stars in low metallicity galaxies produce GRBs. If this is the case we can expect a higher GRB rate in the past, when the fraction of metal-rich galaxies was smaller. Another interesting issue was raised by the detection of the distant GRB 050904 at z = 6.29. The host galaxy of this GRB, while forming stars at a reasonable rate, is not a strong Lyman $\alpha$ emitter, and would have been missed in current Lyman $\alpha$ surveys. This observation emphasizes the ability of GRBs to reveal star forming galaxies which cannot be detected by other means.

\underline{Star formation.} GRBs provide a unique opportunity to trace the formation of massive stars in the early Universe. In particular, they offer the unique opportunity to detect individually (when they explode) the earliest massive stars, the so-called Population III. Massive stars have very short lives (a few millions years), so their birth and their final explosion can be considered as simultaneous in comparison with the age of the Universe. Since long GRBs are detectable at high redshifts, they provide an excellent tracer of the birth rate of massive stars.

\underline{Re-ionization of the intergalactic medium.} UV radiation in the rest frame of a source is efficiently absorbed by neutral hydrogen if this one represents more than 10$^{-3}$ of the total hydrogen. The shape of UV absorption by neutral hydrogen in the vicinity of GRBs at z $\geq$ 6, which is in the near infrared (NIR) region of the spectrum in the rest frame of the observer, has the potential to measure the fraction of neutral hydrogen, providing crucial insight into the history of the re-ionization. The detection of GRB 090423 by {\it Swift} showed that GRBs do indeed exist at redshifts z $>$ 8.

\underline{Cosmological parameters.} Characteristics of GRB spectrum and optical light-curve can be used to standardize the GRB luminosity (for more details, see the chapter by Atteia \& Boer), as it has been already done with type Ia supernovae. Independently measuring the luminosity distance and the redshift provides strong constraints on the geometry of the Universe and on the cosmological parameters. GRBs will provide constraints which are complementary of type Ia supernovae, because they span a different range of redshifts.

\subsection{GRBs and fundamental physics}
\underline{Origin of ultra high-energy cosmic rays.} Theoretical models predict that acceleration processes at work in GRBs could contribute to ultra-high energy cosmic rays. These processes are to be investigated in detail with {\it Fermi}. Unfortunately,  {\it Fermi} cannot provide precise positions for the majority of the GRBs that it detects in the high-energy gamma-ray domain. In this context, external simultaneous GRB detections will play an important role by providing precise localizations of some of the GRBs detected by  {\it Fermi}, allowing measurements of their redshifts and of their energy release.

\underline{Lorentz invariance.} Some quantum-gravity theories suggest that the Lorentz invariance principle is broken in such a way that high-energy photons have a velocity which is very slightly smaller than the velocity of low-energy ones. GRBs offer a powerful tool to test Lorentz invariance by comparing the spectral lags between nearby and distant GRBs. The search for this effect will ultimately require a large number of GRBs with known redshifts and finely sampled light-curves with high signal-to-noise ratio over a broad range of energies.

\underline{Non-photonic messengers.} The times and positions of short GRBs improve the noise rejection capability of large gravitational wave detectors (as e.g. VIRGO and LIGO) which will search gravitational signals coincident with short GRBs. GRBs rank also among the best astrophysical sites able to emit very high-energy neutrinos that will be collected with large area neutrino detectors like IceCube or KM3Net.

\subsection{Non-GRB science}
In any space mission such as {\it SVOM} dedicated to GRB studies, GRB observations occupy less than half of the observing time. The broad wavelength coverage of such a mission allows performing excellent science on selected targets during the remaining of the time. Soft gamma-ray repeaters (SGRs) are one of the most appealing non-GRB targets. SGRs are highly magnetized neutron stars (magnetars), known to repeatedly produce short bursts of soft gamma rays. Only few SGRs have already been identified in the Galaxy and the LMC and it is of prime interest to enrich this small class of remarkable sources. Every 20 years or so, SGRs also emit giant flares which are so bright that they can be detected out to 100 Mpc, well beyond the Virgo cluster of galaxies. It remains to determine the relative rate of giant SGR flares and genuine short GRBs due to merger of binary compact objects. Other interesting non-GRB targets include known astronomical transient sources that will benefit from simultaneous multi-wavelength observations (as e.g. galactic X-ray binaries sources and Active Galactic Nucleis), as well as transient atmospheric phenomena, such as sprites in the upper atmosphere of the Earth.

\section{Scientific requirements of the {\it SVOM} mission}
The scientific rationale, outlined in the previous section, leads to qualitative and quantitative specifications that are detailed below in view of the primary tasks that {\it SVOM} has to fulfill (GRB detection, observation of the prompt GRB emission, near real-time measurement of specific GRB parameters, prompt dissemination of specific GRB parameters, Follow-up of detected GRBs).

\subsection{GRB detection}
At the scale of the Universe, normal (long) GRBs are very rare events, with a rate of about 33$\pm$11 (h$_{65})^3$ Gpc$^{-3}$ yr$^{-1}$ \cite{Guetta}. Collecting a significant amount of GRBs requires long observing time and a detector featuring a wide field-of-view (FOV). As it will be stressed below, GRBs would have to be located onto the celestial sphere with a precision better than 10', thus the choice of possible gamma-ray detector has to be restricted to coded aperture devices, a kind of instrument whose FOV cannot markedly exceed 2 sr without jeopardizing the source location accuracy.

In what follows, it is understood that when detecting a given GRB, the {\it SVOM} detection device will produce a trigger signal at a given time, hereafter designed as the trigger time T$_0$. Note that such a trigger time T$_0$ may occur later than the actual start time of the event. The {\it SVOM} detection device has to be designed to produce such a trigger signal both in the case of long and short duration events. The {\it SVOM} detection device has also to efficiently operate in the X-ray band, down to an energy threshold of a few keV, in order to produce a trigger signal both in the case of high-redshift GRBs whose prompt emission is shifted toward the X-ray band, and in the case of GRBs particularly rich in X-rays.

The {\it SVOM} observing program should be adjusted to allow, in more than 75\% of the cases, prompt follow-up observations with the largest ground telescopes (8 m class). This latter requirement is imposed by the fact that the scientific return of any GRB mission depends chiefly of the amount of data that can be collected by the largest ground telescopes when observing as quickly as possible the GRB afterglow emission and, later on, the GRB host galaxy.

\subsection{Observation of the prompt GRB emission}
To fulfill the two main mission science objectives, related either to the use of GRBs in cosmology and to the understanding of the GRB phenomenon itself, the {\it SVOM} mission requires multi-wavelength observations of the GRB prompt and early afterglow emission. It is also of prime importance to collect multi-wavelength data on the possible emission that may appear before the main GRB prompt emission in order to grasp the first photon signal possibly released by the GRB central engine. In a similar way, it is mandatory to collect multi-wavelength data after the main GRB prompt emission to observe the transition between the prompt emission and the early afterglow, and to insure a deep follow-up of the afterglow.

In hard X-rays and gamma rays, the GRBs detected by {\it SVOM} have to be observed 5 min before and 10 min after the trigger time T$_0$ in the widest energy range: from a hard X-ray lower threshold of 4 keV to a gamma-ray upper-threshold of about 5 MeV. Observations in the X-ray band are of prime importance to detect a possible X-ray precursor signal while observations from few keV to few MeV are mandatory to determine as accurately as possible E$_{peak}$, the GRB peak energy. Simultaneous observation in the visible of a fraction ($>$ 20\%) of {\it SVOM} GRBs down to a limit magnitude M$_V$ = 15 could provide the broadband spectrum of the prompt emission, which is almost unexplored now. 

\subsection{Near real-time measurements of specific GRB parameters}
It has been already stressed that the scientific return of any GRB mission depends chiefly upon the amount of data that can be collected by the largest ground telescopes observing the GRB afterglow as quickly as possible. It is then mandatory to establish for each GRB detected by {\it SVOM} a chain of measurements starting from the space platform down to the Earth. This series of measurements has to involve (i) measurements to be performed by the {\it SVOM} GRB detection device itself, (ii) measurements performed by on-board and ground-based follow-up telescopes triggered by the {\it SVOM} detection, (iii) measurements performed by large ground based telescopes triggered by the follow-up telescopes.

The detection device must be designed in such a way that the determination of the position of detected GRBs is done in few seconds, with relative accuracy (in the detector reference frame) better than 10'. This accuracy is required for the feasibility of follow-up narrow FOV telescopes on-board and on the ground. The experience gained with the {\it Integral} Burst Alert System shows that this delay is to be shorter than 10 s for most GRBs.

While the determination of the burst coordinates is the most critical task, entirely relying on the GRB detection device, it is also essential to compute important properties of the prompt emission in near real time because these properties may impact the follow-up strategy. A summary of the burst consisting of the count rates in selected energy bands will be used to compute essential burst properties (duration, fluence in few energy bands, peak flux, E$_{peak}$, ...). These parameters can be determined in-flight, or more realistically on the ground from data sent to the ground via a near real-time communication link.

\subsection{Prompt dissemination of specific GRB parameters}
As soon as the GRB celestial coordinates have been computed at the level of the GRB detection device, the most accurate measurement is to be made available as quickly as possible (in less than 1 min after T$_{0}$) to ground-based telescopes. Such a short time delay of 1 min is imposed by the rapid decline of the GRB afterglow emission. The prompt knowledge of the GRB celestial coordinates is indeed of prime importance for follow-up telescopes, which can then observe the brightest possible afterglow emission, allowing a more secure identification of the GRB afterglow emission within the field stars. Furthermore, the follow-up telescopes are then in a position to sharpen the measurement of the GRB celestial coordinates to accuracy better than 1'' and to promptly trigger the large ground based telescopes that will undertake useful observations of the still bright afterglow emission. Note that the accuracy of 1'' is precisely the one needed by the powerful spectrographs that are mounted at the focal plane of the largest telescopes to perform the most accurate spectroscopic measurements of the GRB afterglow emission.

\subsection{Follow-up of {\it SVOM} GRBs}
Follow-up telescopes play a decisive role in the scientific return of the {\it SVOM} mission. First, they are the indispensable link between the high-energy GRB detectors which promptly localize GRBs with arcmin precision and the large ground based telescopes which need arcsec positions. Second, they provide unique science on the spectro-temporal evolution of the burst during stages where complex physics is at work: the prompt emission (for the longest GRBs), the transition between the prompt emission and the afterglow, and the early afterglow. The {\it SVOM} mission includes four dedicated follow-up narrow FOV telescopes detailed in the next section, two on-board and two on the ground.

The on-board narrow FOV telescopes will observe nearly 90\% of {\it SVOM} GRBs in visible and X-ray bands. The position of the afterglow (when it is detected) will be refined to several arcsec with a telescope operating in the soft X-ray band and to sub-arcsec with a telescope operating in the visible band. The prompt observation of the GRB field with the visible telescope will permit the exploration of ``dark GRBs'', possibly reducing these events to two classes: high-redshift GRBs and highly extinct GRBs. Both classes are extremely interesting events, which can be studied in X-rays and, from the ground, in the near infrared.

If correctly disposed on the Earth, the ground based telescopes are in a position to observe the early afterglow of the {\it SVOM} GRBs in more than 40\% of the cases. They are needed to measure the celestial coordinates of the GRB afterglow with accuracy better than 0.5'' and to provide an estimate of its photometric redshift in less than 5 min after the trigger time T$_{0}$. This latter requirement is connected with the need to promptly pinpoint the most distant GRBs, i.e. those that are of prime importance for the scientific return of {\it SVOM} as far as cosmology is concerned. The {\it SVOM} alert system will then be in a position to trigger follow-up observations with the largest ground based telescopes with the most convincing set of data.

\section{A mission fulfilling the science requirements}
\subsection{Scientific instrumentation}
\begin{figure}[htbp]
\begin{center}
\includegraphics[width=.8\textwidth]{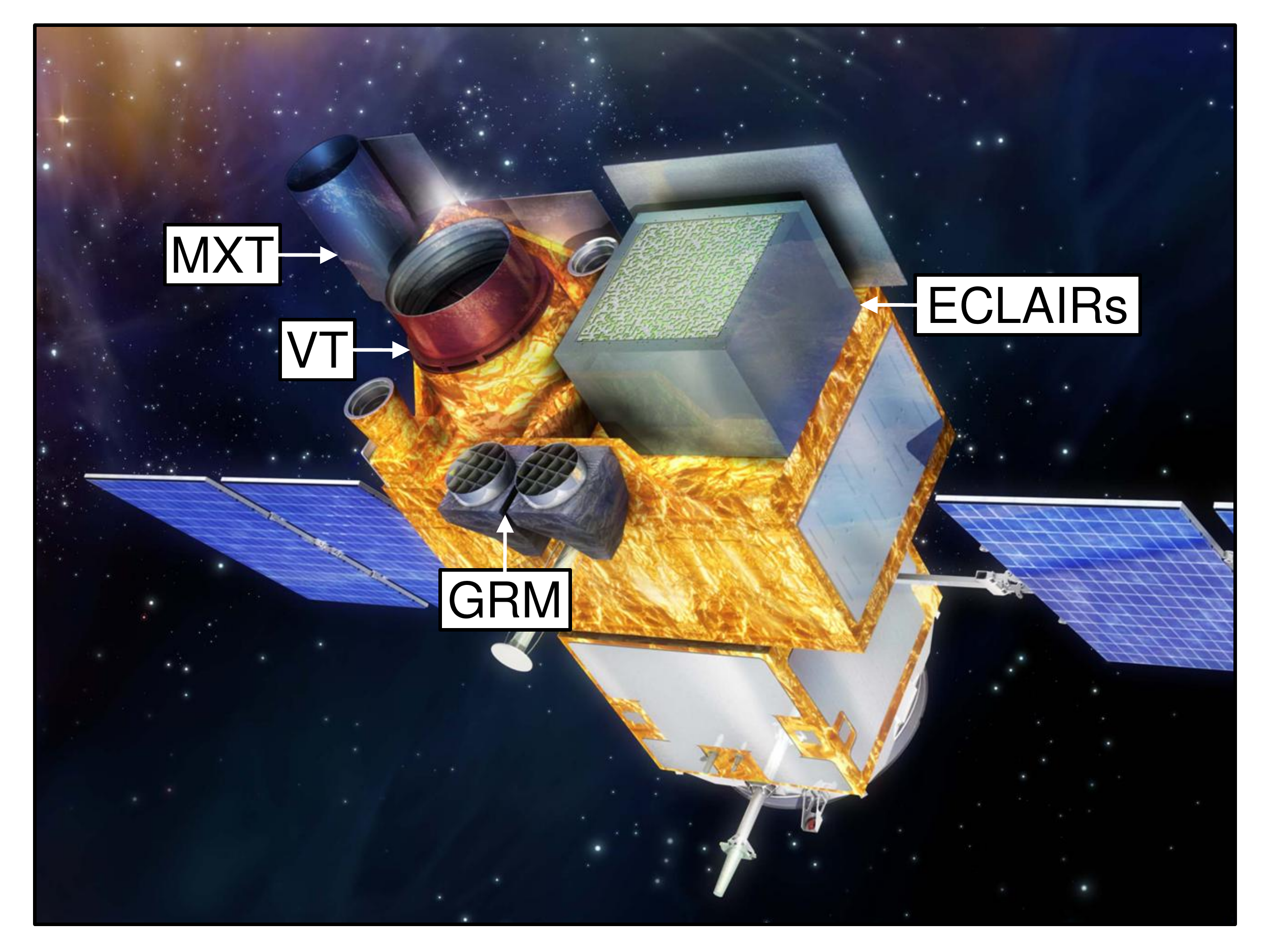}
\caption{Sketch of a possible implementation of the {\it SVOM} payload (copyright: CNES/Oliver Sattler, 2008).}
\label{fig1}
\end{center}
\end{figure}

Figure \ref{fig1}  shows a sketch of the {\it SVOM} satellite carrying a scientific payload composed of four instruments:
\begin{description}
\item[ ] (i) ECLAIRs, a wide-field coded-mask telescope operating in the hard X-ray and soft gamma-ray band for GRB real-time localizations to few arcmin,
\item[ ] (ii) GRM, a non-imaging spectro-photometer for monitoring the FOV of ECLAIRs in the gamma-ray energy range,
\item[ ] (iii) MXT, a narrow FOV telescope for the study of the GRB afterglow in the soft X-ray band,
\item[ ] (iv) VT, a narrow FOV telescope for the study of the GRB afterglow in the visible band.
\end{description}

The space instrumentation is complemented by ground based instruments:
\begin{description}
\item[ ] (v) GWAC, an array of cameras for monitoring the FOV of ECLAIRs in the visible,
\item[ ] (vi) GFTs, two robotic telescopes for the study of the GRB afterglow in the visible and NIR bands.
\end{description}

\smallskip
A description of these instruments is given in section $5$.

\subsection{Satellite orbit}
One important constraint on the satellite orbit is the reduction of the background of high-energy instruments. Like the photons coming from celestial sources, the cosmic photons enter the telescope aperture to give direct interactions. They can also be transmitted through the shields defining the FOV thanks to the high penetrating power of the photons. While photons are the main source of background below 100 keV, charged particles can also enter the FOV of the instruments, producing direct ionization, or they can interact with the detectors or the surrounding matter through nuclear interactions, to generate radioactive species highly increasing the local photon intensity environment with a variable component. Such an activation of the instruments in space is the main source of background above 100 keV. In order to lower the background of the high-energy instruments, they have to be operated in the cleanest possible particle environment.

The Earth magnetosphere is known to trap low and medium energy particles in particular zones (radiation belts). The radiation belts correspond to sites in which energetic particles are highly concentrated, and then, very inappropriate for high energy observations. The zones situated in the latitude intervals +60$^{\circ}$ to +90$^{\circ}$ and  -60$^{\circ}$ to  -90$^{\circ}$ are within the polar caps, thus directly exposed to the primary radiations and to the particle precipitation events. The zones situated in latitudes between -30$^{\circ}$ and +30$^{\circ}$ are favored by a very good protection from ground altitudes up to 650 km.

The magnetic dipole axis of the Earth is not passing through the center of Earth. This little geometrical default is generating inhomogeneities of the magnetic field lines, called the ``South Atlantic Anomaly'' (SAA). This anomaly is equivalent to a hole in the magnetic protection shield of the Earth, in which particles can be injected down to very low altitudes. The shape of this anomaly is roughly a cone pointed in the direction of the ground. This peculiar geometry is such that the zones of particle injection appear to be greater at higher altitudes. In this context, orbits with altitude $h$ $\le$ 600 km and inclination $i$ $\le$ 30$^{\circ}$ would minimize the SAA impact on the trigger camera efficiency. Another constraint on the satellite orbit comes from the alert system as described in section $4.4$ which includes a network of simple receiving stations on the ground. A low inclination of the orbit will result in a reasonable number of the receiving stations.

\subsection{Pointing strategy}
As GRBs are randomly distributed on the sky and in time, the basic philosophy to detect them relies on the implementation of large FOV instruments, which have greater chances to catch random events. The accommodation of a multi-wavelength instrument set, on-board a single platform, (observing from the visible band up to the gamma rays), has to face the different issues expressed by each of them. The Sun, the Earth, and the Moon are all sources of trouble for sensitive instruments operating in the visible, inducing, at minimum, sensitivity losses, and at maximum, destructive damages. The high-energy instruments are less sensitive to these sources, but they are occulted by the Earth, thus decreasing the observing time.

The need of a fine spectroscopy (done by large ground based  telescopes) imposes the immediate visibility from the ground of the detected events. The most powerful instruments on ground are situated close to the tropics. This has also to be taken into account in the observing strategy by selecting pointing directions favoring the observation of {\it SVOM} GRBs by highly sensitive instruments on the ground (i.e. close to their zenith during their night time). In order to permit follow-up observations by large ground telescopes, the {\it SVOM} instruments point close the anti-solar direction during a large fraction of the orbit. For details, see \cite{Cordier}.

\subsection{The alert system}
The need for fast communications from space to the ground has already been emphasized. The solution adopted for the {\it SVOM} mission is an {\it HETE}-like system involving low rate VHF transmission of essential parameters. The system consists in having a VHF emitter on-board the spacecraft and a network of VHF antennas on the ground, under the pass of the satellite. The minimum delay between the alert emission time from the satellite and the time the message is received on the ground was about 10 s in the case of the {\it HETE}-2 mission. In certain cases, this delay rises up to 1 or 2 min. For a low-inclination orbit up to 30$^{\circ}$, like the one considered for the {\it SVOM} mission, about 38 VHF ground stations are required for getting short alert delays compliant with the scientific objectives.

\section{{\it SVOM} scientific instruments}
\subsection{ECLAIRs}
ECLAIRs is a coded-mask telescope coupled with a real-time data-processing unit. Its detector plane consists of 80$\times$80 CdTe semiconductor detectors. A new generation readout electronics, a careful CdTe detector selection and an optimized hybridization allow a fairly low detection threshold. Table \ref{tab1} provides the main characteristics of the ECLAIRs telescope. For details, see \cite{Mandrou}.

\begin{table}[htdp]
\begin{center}
\begin{tabular}{|l|l|}
\hline
Field of view &	2 sr \\
\hline
Energy range	& 4 keV - 250 keV \\
\hline
Sensitive area	& 1024 cm$^2$ \\ 
\hline
Sensitivity (1 000 s, 5 $\sigma$)	& 30 mCrab \\
\hline
GRB localization accuracy &	$<$ 10' (ECLAIRs reference frame) \\
\hline
Burst detection rate	& 70-90 per year (see section $5.8$) \\
\hline
\end{tabular}
\caption{ECLAIRs characteristics}
\label{tab1}
\end{center}
\end{table}

The ECLAIRs data-processing unit analyzes the telescope data stream in real time to detect and localize GRB occurring within its FOV. It implements two simultaneous triggering algorithms \cite{Schanne}: one based on the detection of excesses in the detector count-rate followed by imaging as trigger confirmation and localization, and one performing imaging on a recurrent time-base (better suited for long, slowly rising GRB). After GRB detection, its position is quickly sent to the ground. The GRB position is also transmitted to the platform for an autonomous satellite slew (within five minutes) in order to bring the GRB in the narrow FOV of the onboard MXT and VT telescopes.

\subsection{GRM}
The Gamma-Ray Monitor (GRM) is composed of two detector modules, each made of NaI/CsI active layers and a collimator for background reduction and FOV restriction to the one of ECLAIRs. The GRM does not provide imaging capability, but extends the {\it SVOM} spectral coverage into the MeV range, such that every {\it SVOM} GRB will have a good ECLAIRs/GRM spectrum. This is very important in order to determine a key GRB parameter, E$_{peak}$. Table  \ref{tab2} provides the main characteristics of the GRM. For details, see \cite{Dong}.

\begin{table}[htdp]
\begin{center}
\begin{tabular}{|l|l|}
\hline
Number of units&	2 \\
\hline
Field of view	&2 sr \\
\hline
Energy range	& 30 keV - 5 MeV \\
\hline
Sensitive area	& 280 cm$^2$ (each unit) \\
\hline
Sensitivity (50-300 keV, 1 s, 5 $\sigma$)	& 0.23 photons cm$^{-2}$ s$^{-1}$ \\
\hline
\end{tabular}
\caption{GRM characteristics}
\label{tab2}
\end{center}
\end{table}

\subsection{MXT}
The Micro-channel X-ray Telescope (MXT) is a focusing X-ray telescope, featuring a micro-channel plate lens inherited from MIXS-T, an instrument to be flown on the ESA Mercury mission Bepi-Colombo \cite{Martindale}. It employs a compact, low-noise, and fast read-out CCD camera. Table \ref{tab3} provides the main characteristics of the MXT.

\begin{table}[htdp]
\begin{center}
\begin{tabular}{|l|l|}
\hline
Field of view	& $>$ 25' \\
\hline
Effective area	& 50 cm$^2$ (at 1 keV) \\
\hline
Focal length     &1 m \\
\hline
Energy range	& 0.3-6 keV \\
\hline
Sensitivity (10 s, 5 $\sigma$) & 	2.5 mCrab \\
\hline
Angular resolution   & 2' (at 1 keV) \\
\hline
GRB localization accuracy	& $<$ 30'' (50\% of the cases)  \\
\hline
\end{tabular}
\caption{MXT characteristics}
\label{tab3}
\end{center}
\end{table}

An accurate localization of the GRB is essential to allow successful follow-up observations. After GRB detection and localization with ECLAIRs, MXT must be able to detect the X-ray emission from the GRB afterglow, and to compute in near real time a first GRB position (which is immediately sent to ground via VHF and to the VT). It must also be possible to derive improved positions as more data are accumulated by MXT, and, if the improved positions are significantly better than the previous ones, they will also be transmitted to ground.

\subsection{VT}
The space-borne Visible Telescope has been designed to fulfill the following requirements: (i) to improve the GRB localizations obtained by the ECLAIRs and MXT to sub-arcsec precision through the observation of the optical afterglow, (ii) to obtain a deep and uniform sample of optical-afterglow light-curves, (iii) to do a primary selection of optically dark GRBs and high-redshift GRB candidates (z $>$ 4). Table  \ref{tab4}  provides the main characteristics of the VT.

\begin{table}[htdp]
\begin{center}
\begin{tabular}{|l|l|}
\hline
Diameter	& 450 mm \\
\hline
Optical design	& Modified Ritchey-Chr\'etien\\
\hline
Field of view	& 21' $\times$ 21'\\
\hline
Spectral bands	& 400-650 nm and 650-950 nm\\
\hline
Detector	& 2 CCD cameras\\
\hline
Angular resolution	& 0.6''\\
\hline
Sensitivity (300 s, 5 $\sigma$)	& M$_V$ = 23\\
\hline
GRB observation rate	& 80\% of {\it SVOM} GRBs\\
\hline
\end{tabular}
\caption{VT characteristics}
\label{tab4}
\end{center}
\end{table}

\subsection{GWAC}
The Ground-based Wide-Angle Camera (GWAC) array permanently observes a fraction of the ECLAIRs FOV, to search for visible emission of more than 20\% of the {\it SVOM} GRB. Table \ref{tab5} provides the main characteristics of the GWAC.

\begin{table}[htdp]
\begin{center}
\begin{tabular}{|l|l|}
\hline
Number of units &	64\\
\hline
Field of view	& $\sim$8000 deg$^2$ ($\sim$130 deg$^2$ per unit)\\
\hline
Spectral band	& 400-950 nm\\
\hline
Detector	& 64 CCD cameras (one per unit)\\
\hline
Sensitivity (10 s, 5 $\sigma$)	& M$_V$ = 16 (new Moon) \\
\hline
GRB observation rate	& 20\% of {\it SVOM} GRBs\\
\hline
\end{tabular}
\caption{GWAC characteristics}
\label{tab5}
\end{center}
\end{table}

\subsection{GFTs}
Two dedicated robotic Ground-based Follow-up Telescopes (GFTs) automatically position their FOV to GRB alert positions. In case of counterpart detection, they improve the GRB localization accuracy to 0.5''. Both telescopes use multi-band optical cameras, and the French GFT has additional near infra-red capabilities. The telescope sites are located in tropical zones, at longitudes separated by 120$^{\circ}$ in order to optimize the GRB follow-up capability. Low significance alerts, considered not reliable enough to be distributed to the whole community, are also followed by those dedicated telescopes. Table \ref{tab6} provides the main characteristics of the two GFTs.

\begin{table}[htdp]
\begin{center}
\begin{tabular}{|l|l|l|}
\hline
	& Chinese GFT	 & French GFT\\
\hline
Spectral band	&400-950 nm	& 400-1700 nm\\
\hline
Field of view	&21' $\times$ 21' 	& 30' $\times$ 30'\\
\hline
Aperture	& 1 000 mm	& 980 mm\\
\hline
Detector	& 3 CCD cameras &	2 CCD / 1 IR camera \\
\hline
Sensitivity (5 $\sigma$)	 & M$_R$ = 21.5 (100 s)	 & M$_J$ = 18 (10 s)\\
\hline
GRB observation rate &	20\% of {\it SVOM} GRBs	& 20\% of {\it SVOM} GRBs\\
\hline
\end{tabular}
\caption{GFTs characteristics}
\label{tab6}
\end{center}
\end{table}

\subsection{{\it SVOM} multi-wavelength capabilities}
\begin{figure}[htbp]
\begin{center}
 \includegraphics[width=.8\textwidth]{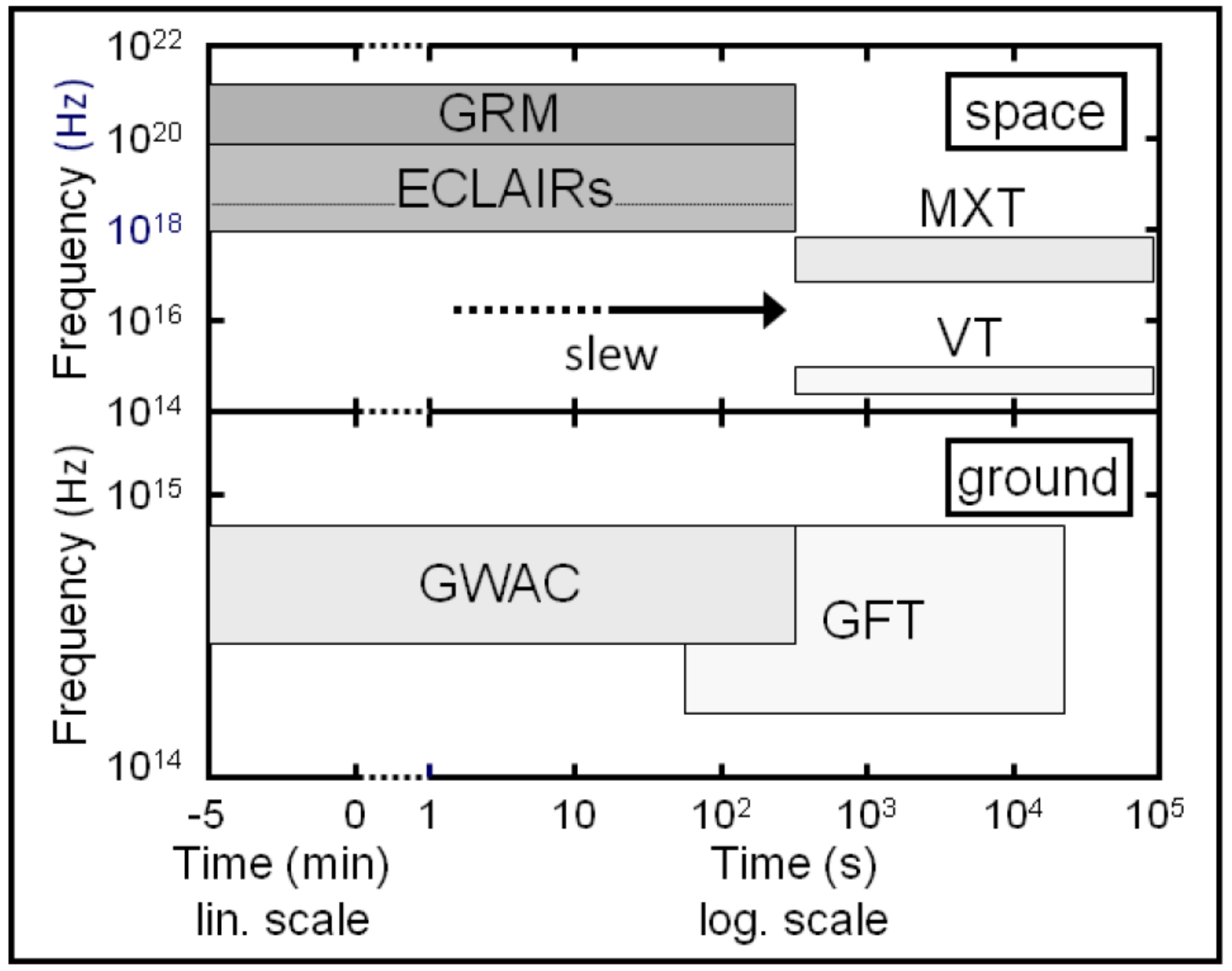}
\caption{Spectral coverage as a function of time of the prompt and afterglow emission of {\it SVOM} GRBs.}
\label{fig2}
\end{center}
\end{figure}

Figure \ref{fig2} illustrates the multi-wavelength coverage of GRB emission (prompt and early afterglow) with {\it SVOM}. For tens of GRBs per year, {\it SVOM} will provide the broadband coverage of the GRB prompt emission (from visible to MeV), the detailed photometry of the afterglow, and often the measure of the redshift. This should lead to completely new constraints in the GRB field.

\subsection{{\it SVOM} anticipated performances}
\begin{figure}[htbp]
\begin{center}
 \includegraphics[width=.8\textwidth]{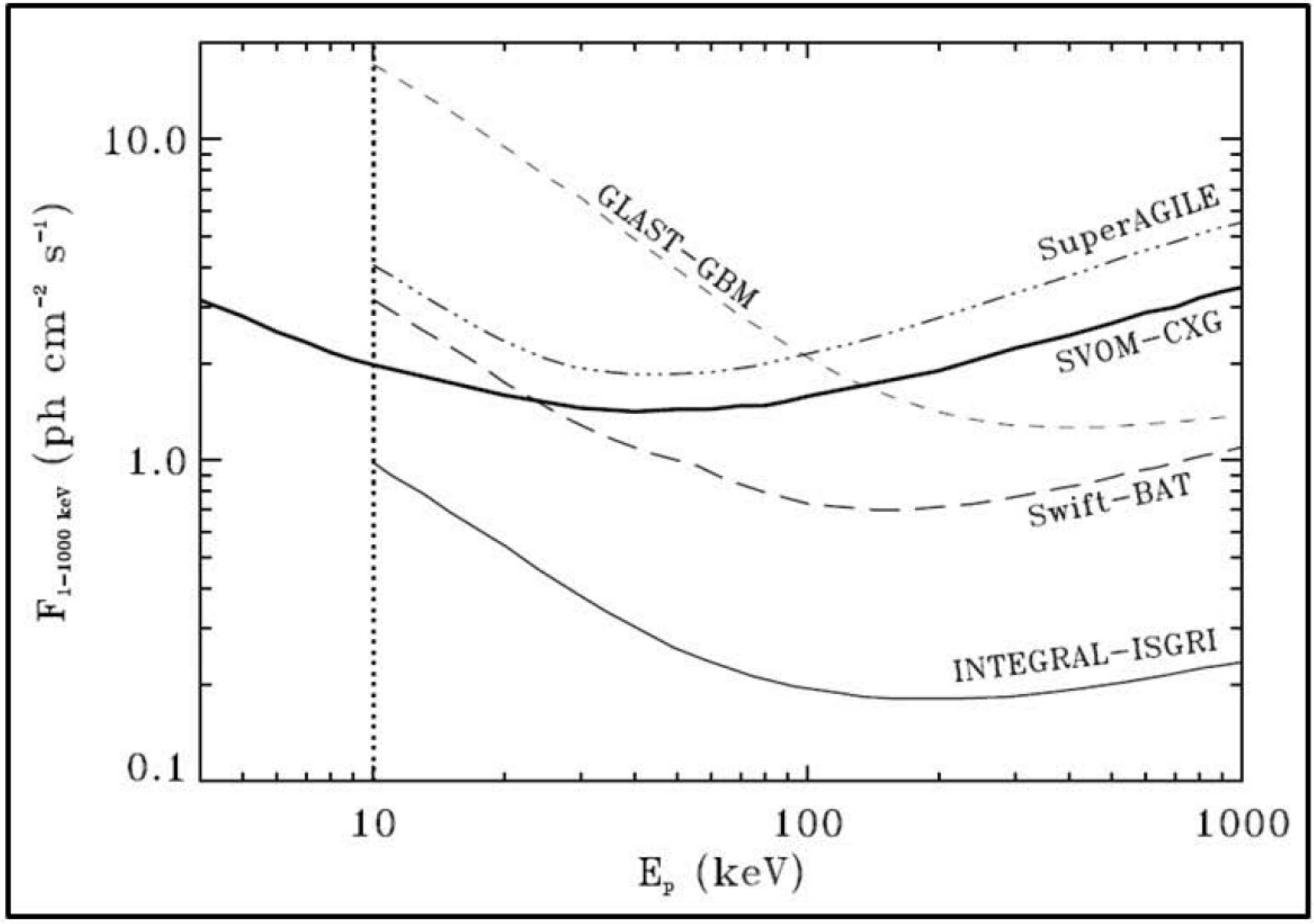}
\caption{GRB detection sensitivity of ECLAIRs in the 1-10$^3$ keV energy range as a function of the GRB peak energy, compared with that of previous missions (from \cite{Godet}).}
\label{fig3}
\end{center}
\end{figure}

Detailed simulations of the number of GRB detections expected with {\it SVOM} have been performed. These simulations rely on the description of a full GRB population characterized by the following quantities: GRB co-moving rate, GRB luminosity function, GRB spectral parameters, durations and variability. They also include the ECLAIRs sensitivity (see Figure \ref{fig3}).

These simulations lead to a GRB detection rate of 70-90 GRBs/yr depending on the assumptions on the GRB population. In particular, it is shown that nearly 20\% of ECLAIRs GRBs could be situated at high redshifts (z $>$ 6). It should also be stressed that owing to the SVOM pointing strategy (see section 4.3), it is anticipated that redshifts could be estimated for 75\% of the GRBs. SVOM would then be in a position to grasp per year more GRBs with redshift measurements than Swift. Recently, {\it Swift} has detected a few nearby GRBs with low luminosities (as e.g. GRB 060218). These events, similar to GRB 980425 at z = 0.0085, suggest the existence of a large population of faint and soft nearby GRBs, usually called sub-luminous GRBs. With its low energy threshold, ECLAIRs will be particularly sensitive to this population. Given the uncertainties in the characteristics and number of such nearby GRBs, the evaluations of GRB rates given above do not take this population into account. 

Noting finally that many short GRBs have been found to be followed by a period of extended emission in the X-ray range, the remarkable sensitivity of ECLAIRs at low energy will permit the localization of short GRBs from their extended emission, even if the brief initial pulse does not contain enough photons to obtain a reliable localization.

\section{Conclusion}
With a launch in the second half of the present decade, {\it SVOM} will most probably be the main provider of accurate GRB positions during its lifetime. In this time frame it is envisioned that the interest in GRBs will remain high, with many important questions actively studied like: the physical mechanisms that launch the relativistic jet, the radiation processes responsible for the prompt emission, the elucidation of the connection between long GRBs and supernovae, and the origin of the short/hard GRBs and of other sub-classes of GRBs, like sub-energetic events and X-ray flashes. Furthermore, the quest for high redshift GRBs will continue since these events constitute valuable tools for the study of the star formation rate, of the metal enrichment in the young Universe, and for the investigation of the re-ionization of the Universe by the first luminous objects at the end of the dark ages. It will also be highly important to study local GRBs (e.g., z $<$ 0.3) which permit accurate studies of the presumably associated supernova and the detailed exploration of their host galaxies and local environment. 

This period will be particularly appropriate for GRB follow-up, with a number of existing facilities still active, like the current networks of small robotic telescopes, and the availability of completely new instrumentation. In this second class we can mention mid-size telescopes (1-2 m), which are now being automated and equipped with visible and NIR cameras. Various larger facilities will be still available, as e.g. the  {\it Fermi} satellite devoted to high-energy gamma-ray observations, {\it X-Shooter} and other second-generation instruments at the ESO-VLT. During this period, numerous new astronomical detectors are also going to be operational: {\it JWST}, the NASA-ESA large space infrared observatory, {\it POLAR}, a GRB polarization experiment onboard the Chinese space laboratory, advanced gravitational radiation detectors ({\it advanced Ligo/Virgo}), large radio telescope arrays ({\it LOFAR}, {\it SKA}), large neutrino detectors ({\it IceCube}, {\it KM3Net}), large Cherenkov telescope array ({\it CTA}) and new ultra high energy cosmic-ray detectors, just to quote the most relevant ones.

A GRB satellite during such a period could enable detection of GRBs in coincidence with various phenomena at other wavelengths or even with gravitational radiation or neutrinos. For example, the leading model candidates for sources of short GRBs involve neutron star mergers. These are also the prime candidate sources for gravitational radiation detectors. A simultaneous detection of a GRB and a gravitational radiation signal would be the ultimate proof of this model and open a new window on the Universe. As such the role of {\it SVOM} will be unique. This excellent instrumental conjunction, combined with the expected progresses of visible ground telescopes and with the relevance of the scientific issues connected with GRB studies warrants a remarkable scientific return for {\it SVOM}.

\end{document}